# GANG-MAM: GAN based enGine for Modifying Android Malware

Renjith G[a], Sonia Laudanna[b], Aji S[a], Corrado Aaron Visaggio[b], Vinod P[c]

[a]*Department of Computer Science, University of Kerala, Trivandrum, Kerala, India*
[b]*Department of Engineering, University of Sannio, Italy*
[c]*Department of Computer Applications, Cochin University of Science and Technology, Kochi, Kerala, India*

**Abstract**

Malware detectors based on machine learning are vulnerable to adversarial attacks. Generative Adversarial Networks (GAN) are architectures based on Neural Networks that could produce successful adversarial samples. The interest towards this technology is quickly growing. In this paper, we propose a system that produces a feature vector for making an Android malware strongly evasive and then modify the malicious program accordingly. Such a system could have a twofold contribution: it could be used to generate datasets to validate systems for detecting GAN based malware and to enlarge the training and testing dataset for making more robust malware classifiers.

*Email addresses:* renjithg@keralauniversity.ac.in (Renjith G), slaudanna@unisannio.it (Sonia Laudanna), aji@keralauniversity.ac.in (Aji S), visaggio@unisannio.it (Corrado Aaron Visaggio), vinod.p@cusat.ac.in (Vinod P)

## Required Metadata

## Current code version

| Nr. | Code metadata description | Please fill in this column |
|---|---|---|
| C1 | Current code version | Version 1.0 |
| C2 | Permanent link to code/repository used for this code version | https://github.com/papersubKS |
| C3 | Legal Code License | GPLv2 |
| C4 | Code versioning system used | git |
| C5 | Software code language used | Python 3.6+ |
| C6 | Compilation requirements | Compiled and tested in Ubuntu 18.04 LTS, Tensorflow 2.4.1, Keras 2.4.3, Scikit-learn 0.23.1, Matplotlib 3.2.2, Apktool 2.5.0, QEMU 2.11.1, Android Emulator 30.7.5.0, Android SDK Tools 26.1.1, Android SDK Platform-Tools 29.0.5, AAPT 0.2-27.0.1, Java 1.8 |
| C7 | If available Link to developer documentation/manual | https://github.com/papersubKS |
| C8 | Support email for questions | papersubKS@gmail.com |

Table 1: Code metadata (mandatory)

## 1. Motivation and significance

Generative Adversarial Networks (GAN) [1] were introduced to produce adversarial samples that are able to lead machine learning based classifiers to the wrong decision. The main concept behind a GAN is that two independent Neural Networks, the Generator and the Discriminator, through a competition, create fake samples that are classified as real samples by a classifier. The Generator creates the fake candidate samples, while the discriminator makes a judgement on the fake samples and the real data. On the basis of this judgement, the Generator generates fake data that are more similar to the real ones. After an iterative process, the Generator is able to create fake data indistinguishable from the real data. Recent studies focused on the production of adversarial examples with GAN that deceive malware classifiers into recognizing malicious files as legitimate. MalGAN [2] is a pioneer work on black-box adversarial-example attacks toward Android malware detection. It used a GAN to generate adversarial examples to bypass black-box detectors. Subsequent studies investigated how the GAN can be used to realize effective adversarial samples. E-MalGAN [3] is an evasive adversarial-example attack that misleads in-cloud firewall-equipped Android malware detection systems without requiring any information about the target. GAN was also used to create adversarial samples of Windows malware [4], while in [5] authors used deep Convolution GAN for allowing training with a few data. In [6] a deep learning GAN can create images that appear to be malware samples visualized as an image. Authors in [7] do the same thing but using auxiliary classifier GANs. In this paper, we will introduce a system which allows to modify an existing malware on the basis of a GAN output. The GAN determines which are the most suitable feature vectors for the malware, and then the system modifies the malware accordingly.

GAN generated malware can be used in malware research in two main ways. The first one is to improve the effectiveness of malware detectors by enlarging the training dataset. The second one is to develop methods that can recognize malware produced by GAN, which is indistinguishable from benign samples by classifiers. Considered the increasing interest of the malware analysis community in GAN generated malware, the software that we propose can support researchers to easily build datasets for their experiments.

The main novelty of our system is that it can modify existing Android malware, maintaining its operational integrity. We will make available our software to scientific research in the public repository[1].

## 2. Software description

We present an automated tool aiming at making existing malware strongly evasive, namely GAN Generated malware for Modifying Android Malware (GANG-MAM), based on a GAN engine. The tool consists of two parts: the GANG that produces an evasive features vector and the MAM that accordingly modifies the existing malware, preserving its malicious behaviour.
For a given malware $m$, whose the corresponding feature vector is $V_m$, the GANG produces a new feature vector $V_m^o$, that will let the malicious program be recognized as benign by a malware classifier. Of course, the malware $m$ must properly change in order to show the new features vector. This task is accomplished by the MAM engine, which transforms the malware $m$ in $m^o$ by updating the corresponding features. The block diagram of the tool is shown in Figure 1.

---
[1]https://github.com/papersubKS

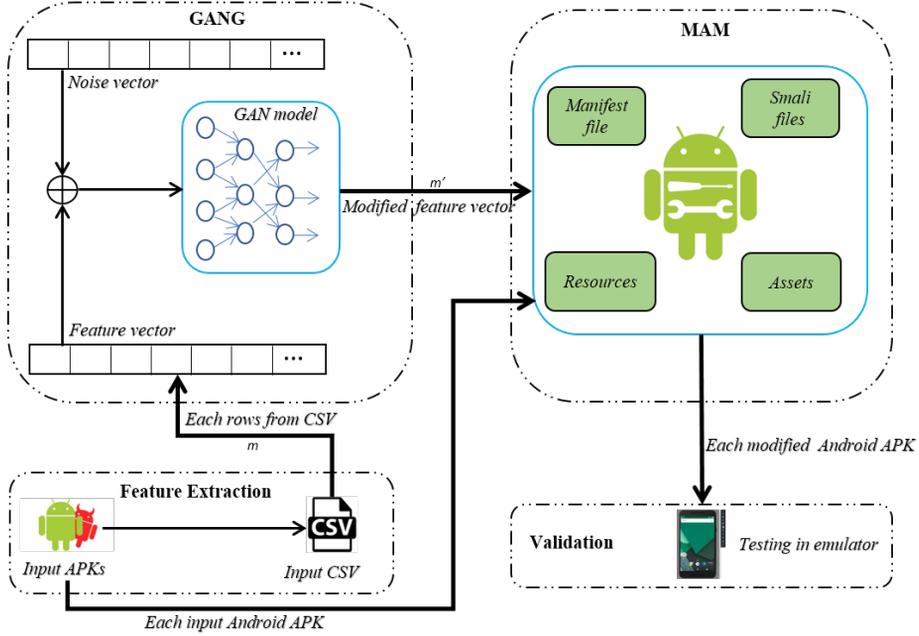

Figure 1: The high-level Architecture of GANG-MAM

Therefore, the challenge is that the modification of the feature must not compromise malware's original functions, and the perturbation to the feature space can be practically implemented in the Android PacKage (APK), meaning that the modification can be realized in the program code of an unpacked malware and can also be repacked/rebuilt into the original malicious APK. The MAM modifier works on the input APK file based on an input CSV file that contains the features the existing APK, generated by the GANG, must show. Once the modifications are made, the tool repackages the APK, signs it and generates the output APK. The underlying features of the APK can be easily modified without accessing its original Java/Kotlin source code. MAM is designed to handle any number of input APKs with the capability to modify different types of features in each APK. The tool ensures a hassle-free process to specify the features to be modified via an input feature definition CSV file. Additionally, the tool can also insert smali equivalents of the newly added activities, services, broadcast receivers and content providers. Re-engineering the contents of an APK is made easier as the tool automates all the repetitive processes like APK assembling/disassembling, feature modification, building, signing, and testing. Table 2 shows an example of the captured features.

| Feature | Value |
|---|---|
| Permission | permission.ACCESS_NETWORK_STATE |
|  | permission.ACCESS_WIFI_STATE |
|  | permission.READ_PHONE_STATE |
|  | permission.INTERNET |
| Intents Action | action.FTPSERVER_STOPPED |
|  | action.DOCUMENTSPROVIDER |
|  | action.FTPSERVER_STARTED |
|  | action.UPGRADE_ALL |
| Service | TapContextService |
|  | QuickAccessibilityService |
|  | PluginPitService |
|  | Push_BroadCast_Service |
| Intents Category | category.MULTIWINDOW_LAUNCHER |
|  | category.DEFAULT |
|  | category.BROWSABLE |
|  | category.LEANBACK_LAUNCHER |

Table 2: Example of Static Features

*2.1. Software Architecture*

The detailed software architecture of GANG-MAM is presented in Figure 2. GANG-MAM accepts any number of Android APKs and generates modified Android APKs as output. GANG-MAM has mainly four functions such as feature extraction, GANG engine, MAM engine and validation.

*2.1.1. Feature Extraction*

The feature extraction phase includes dissembling of the input Android APKs and extraction of feature vectors from it. The features considered are permissions, activities, services, receivers, providers, actions and categories, as shown in Table 2. Apktool is used to dismantle each input APKs selected. The actual feature extraction is done on the dismantled APKs. The feature vectors are represented in a CSV format. Each APKs is selected, and a hash is computed. The column entries in the CSV file are the SHA256 hash value for the APKs and all the static features in the APK. In each row, corresponding to each hash value, the feature definitions are presented. Each attribute corresponding to APKs is represented as 1's and 0's, where 1 denotes the presence of an attribute, conversely 0 signify the absence of the feature. While creating new a variant, if elements of the App feature vector are 1, the corresponding feature is added; the addition of extraneous features while retaining the older ones preserves the functionality of an application.

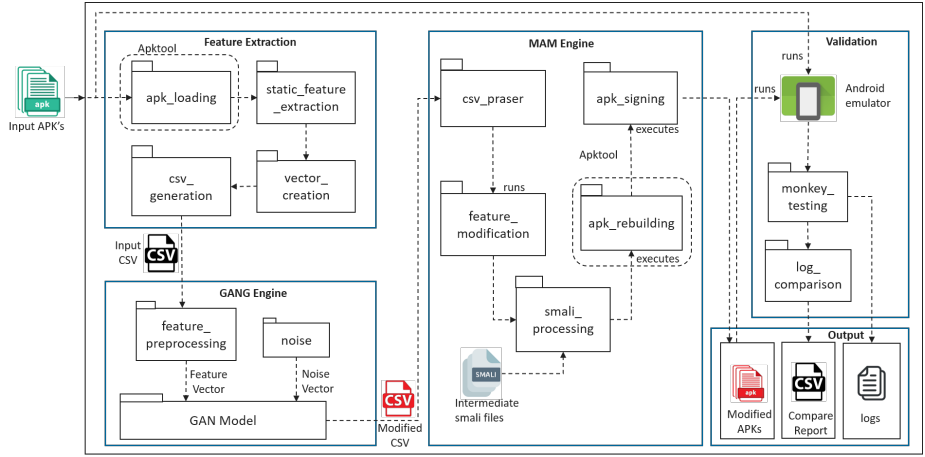

Figure 2: Detailed Architecture of GANG-MAM

### 2.1.2. GANG Engine

GANG engine accepts the input feature vector and generates the modified feature vector. A pretrained GAN model is used to generate the modified feature vector.

### 2.1.3. MAM Engine

MAM engine accepts the modified feature vector CSV file with the updated feature details for each APK. MAM will manage any number of features for all the input APKs. MAM also processes the input intermediate smali files, which facilitate the tool to insert smali equivalents of the newly added static features like activities, services, broadcast receivers and content providers. The inputs are validated before they are used for further processing. MAM selects each input APKs and calculates its hash. The modified feature vector CSV file is parsed to find the corresponding entry of the APK in the CSV file.

Apktool is used to disassemble and reassemble Android applications. This contains the internals of the APK file such as manifest file, assets, resources, intermediate smali files. The features that must be modified for this APK are read from memory and applied to the disassembled APK contents in their corresponding locations. The tool will insert intermediate smali files corresponding to those static features that require run time modifications such as activities, services, broadcast receivers and content providers. Once all the features have been modified as per the feature modification CSV file, Apktool is used to repackage the modified contents such as assets, manifests, resources, smali files and then generates the output APKs and then signs it.

MAM engine can work independently from the GANG engine as well. The input APKs and the corresponding feature vectors to be applied to the APKs should be provided.

*2.1.4. Validation*

The output APKs are validated using the Android emulator environment. Each of the input APKs is installed, launched, executed using monkey [2], uninstalled and cleaned up. In the same way, each of the output modified APKs are installed, launched, executed using monkey, uninstalled and cleaned up. Both the execution logs are stored and compared for any difference in the functional execution flow. This ensures that the modified APKs are functional and behave like input APKs.

The entire GANG-MAM process is fully automatic and repeated till all the input APKs are processed and output APKs are generated and validated.

*2.1.5. Implementation*

GANG-MAM is fully developed using python 3.6+. Apktool 2.5.0 [8] is used to assemble and disassemble the Android APKs. The GANG engine is developed using Tensorflow 2.4.1 [9], Keras 2.4.3 [10] and Scikit-learn 0.23.1 [11]. The Android APK signed using two command-line utilities keytool and jarsigner [12]. The experimental evaluation is conducted on MSI GF63 (Intel Core i7 processor, 9th generation, 16GB memory, NVIDIA GeForce GTX 1650 With Max-Q design, 4GB GDDR6) laptop. The operating system used is Ubuntu 18.04 with virtualisation support. The input and modified APKs are tested and validated using the latest Android 12.0 emulator, and monkey [13] environment.

*2.1.6. Stability Results*

The GANG-MAM tool generates the modified android applications from the input malware applications according to the defined feature vector changes. The feature vector changes can be done using the GANG engine, or we can feed it externally. We have used Androzoo [14] malware repository and downloaded 500 Android applications (Eg: malware) for our stability testing. The tool could successfully generate an input feature vector for these 500 Android applications. GANG engine could modify the feature vectors for these 500 Android applications. MAM engine successfully created 500 modified Android applications at the output. During our validation step, it was observed that almost 492 Android applications ensure operational integrity and were successfully tested in the Android emulator environment with monkey. We have traced the reason for the failure of 8 Android applications and found that these applications require login credentials to run in the Android emulator environment. Table 3 represents the execution details of 10 Android APKs. The execution logs of the Android APK before and after modification are stored in two separate files. The Android APK is installed and launched using 'adb' utility. The installed APK is executed in the Android emulator environment. Android 'monkey' is used to inject the events (with the same seed) and control the application during runtime. The execution logs of original and modified applications are compared using Linux 'diff' utility to compute the difference in the log files.

---
[2]https://developer.android.com/studio/test/monkey

| # | APK names | Before | After | Diff |
|---|---|---|---|---|
| 1 | puzzles.legogames9.legobatman9.apk | 139 | 139 | 1 |
| 2 | com.tianer.cloudcharge.apk | 141 | 141 | 1 |
| 3 | com.thevotinggame.thevotinggame.apk | 35 | 35 | 1 |
| 4 | com.cocoa.cocoa_17875_5715f29.apk | 35 | 34 | 3 |
| 5 | com.landlordvision.mobileapk | 141 | 141 | 1 |
| 6 | tools.app.volume.super.loud.apk | 153 | 153 | 1 |
| 7 | com.robotobia.hdstockwallpapers.apk | 139 | 140 | 2 |
| 8 | com.appybuilder.asker88kudus.M0.apk | 140 | 140 | 1 |
| 9 | com.stmvideo.webtv.radiorevivert.apk | 140 | 141 | 2 |
| 10 | ru.indieproductions.survivalguidef.apk | 141 | 139 | 2 |

Table 3: APK operational integrity (before and after modification)

The differences in 1 to 3 lines are observed while comparing the execution logs. We manually investigated those lines and found that the differences are either related to timestamp information or dynamic input data. In both cases, the operational and functional flow for the Android APK before and after modification is the same.

2.2. Software Functionalities & Usage

The major highlights of this tool are: this is the first tool that generates adversarial Android malware samples; this tool supports end to end automation; this tool has the ability to handle any number of APKs; Also, it supports features to be modified via a separate CSV file that means, modification engine (MAM) can also work independently from the GANG engine; this tool has the ability to handle an unlimited number of static features; besides our tool is independent of the underlying Java/Kotlin source code; easy to integrate the smali for the necessary features; this tool can be extended to automatically modify dynamic features in the APKs.

The command-line options and usage are detailed below:

```
$ ./run.sh -h
```

- -e  Name of the emulator
- -c  Clean all output folder contents created earlier
- -n  Path of the feature vector file to run in No GANG mode
- -v  Print current tool version
- -h  help message

The emulator to be used for testing the modified APKS can be passed to the tool using the 'e' flag. The tool checks for the name of the emulator in the available emulator list, and if found, will use it for testing the APKS. For example:

```
$ ./run.sh -e Nexus_4a
```

To run this tool without feature vector generation (GANG), use the '-n' option. The feature vector to be used should be passed as an argument with this option.

```
$ ./run.sh -n /home/user/feature_vector.csv
```

*2.3. Sample code snippets analysis (optional)*

## 3. Illustrative Examples

A video is available in the Git repository of GANG-MAM project.

## 4. Impact

The proposed software can support the research about GAN produced malware in two directions. The first one is that it can help to enlarge the experimental dataset with adversarial samples. One of the problems of malware analysis research is to collect many instances which must be enough recent properly validated and labelled. This brings about that researchers can not often have a dataset huge enough for obtaining strong evidence from their experiments. The second one is to support the study of methods for contrasting GAN produced malware. GAN is a tool that can be used to produce very evasive malware. Currently, the body of knowledge is missing of methods that can successfully tackle GAN produced malware, while it is an imminent threat.

## 5. Conclusions

We realized GANG-MAM, a tool for modifying existing Android malware with GAN produced features vectors. Due to the large diffusion of classifiers for detecting and analysing malware, many researchers are investigating adversarial attacks and how to protect from them. Since GAN represents an effective method for generating strongly evasive malware, we expect that the tool we propose will be largely used by both research but also industry community. The tool can have a two-fold application: testing the robustness and effectiveness of existing malware classifiers, by finding blind spots in the decision models of classifiers or by enlarging the training data-sets; providing data-sets for testing new methods aiming at recognizing malware produced with GANs.